\documentstyle[aps,pre,preprint,eqsecnum]{revtex}
\tightenlines
\draft

\begin{document}
\title{{\bf Kinetic Theory and Hydrodynamics for Rapid Granular Flow - A
Perspective}}
\author{James W. Dufty}
\address{Department of Physics, University of Florida, Gainesville, FL\\
32611, USA}
\date{\today }
\maketitle

\begin{abstract}
These are notes prepared for presentation at the workshop ''Challenges in
Granular Matter'' at the Abdus Salam Institute for Theoretical Physics,
Trieste, August 2001. Revisions and figures will be added at a later date.

Many features of real granular fluids under rapid flow are replicated by a
system of smooth hard spheres with inelastic collisions. For such a system,
it is tempting to apply standard methods of kinetic theory and hydrodynamics
to calculate properties of interest. The domain of validity for such methods
is {\em a priori} uncertain due to the inelasticity, but recent systemmatic
studies continue to support the utility of kinetic theory and hydrodynamics
as both qualitative and quantitative descriptions for\ many physical states.
The current status of kinetic theory and hydrodynamics is reviewed and
interpreted, with the optimistic conclusion that much existing phenomenology
can be placed on firm grounds for further understanding of complex states
observed in rapid granular flow.
\end{abstract}

\section{Introduction}

Granular media in rapid, dilute flow exhibit a surprising similarity to
ordinary fluids and the utility of a hydrodynamic description for such
conditions has been recognized for many years \cite{haff83}. This
phenomenology has come under scrutiny in recent years with questions about
the domain of its validity and the associated constitutive equations
appearing in the hydrodynamic equations \cite{kadanoff99,goldhirsh01}.
Answers to such questions can be found in a more fundamental microscopic
description where the tools of nonequilibrium statistical mechanics are
available for critical analysis. An intermediate mesoscopic description
between statistical mechanics and hydrodynamics is that of kinetic theory,
whose applicability to granular matter also poses questions. Attention \
here is restricted to an idealized microscopic model system that captures
only the primary feature of grains: inelastic collisions. The system is
composed of smooth, hard spheres whose energy loss on binary collisions is
characterized by a restitution coefficient $0<\alpha \leq 1$. This single
parameter distinguishes the ideal granular fluid $\left( \alpha <1\right) $
from the ideal normal fluid $\left( \alpha =1\right) $. For normal fluids,
extensive studies during the past thirty years confirm the validity of a
wide range of many-body methods to analyze states near and far from
equilibrium. These studies include applicability of kinetic theory
(Boltzmann, Enskog, and corrections due to correlated collisions), and on
longer space and time scales, a hydrodynamic description. The foundations
for kinetic theory and hydrodynamics, including practical calculation of
properties such as the equation of state and transport coefficients, is most
complete for the elastic hard sphere system.

How much of this body of knowledge can be translated to rapid flow granular
media using the same system with the only change being $\alpha <1$? At the
most fundamental level of Newton's equations there is little evident change
for the fluidized state; the microstate of the system at time $t$ is
completely characterized by the positions and velocities of all spheres. The
sequence of free streaming and instantaneous velocity changes on binary
collisions determines the evolution uniquely for given initial conditions at 
$t^{\prime }<t$. Only the binary collision rule changes, such that the
relative velocity ${\bf g}_{ij}={\bf v}_{i}-{\bf v}_{j}$ for particles $i$
and $j$ after the collision is given by%
\begin{equation}
\widetilde{{\bf g}}_{ij}={\bf g}_{ij}-(1+\alpha )\widehat{\text{\boldmath$%
\sigma $}}({\bf g}_{ij}\cdot \widehat{\text{\boldmath$\sigma $}}).
\label{1.1}
\end{equation}%
where $\widehat{\sigma }$ is a unit vector along the line from particle $j$
to $i$. This is the usual rule for reflection of relative velocities except
that the magnitude of the normal component $\left| \widetilde{{\bf g}}%
_{ij}\cdot \widehat{\text{\boldmath$\sigma $}}\right| $ after the collision
has been decreased by a factor $\alpha $ to $\alpha \left| {\bf g}_{ij}\cdot 
\widehat{\text{\boldmath$\sigma $}}\right| $. Thus, the standard event
driven molecular dynamics (MD) simulation method developed for normal fluids
can be applied directly to granular fluids with this small change in the
collision rule\footnote{%
The phenomenon of ''inelastic collapse'' can stop the MD simulation due to
condensation of small arrays \ of particles with rapid collisions at high
density and strong dissipation \cite{kadanoff99}. This is an artifact of the
instantaneous hard sphere collision that appears to have no macroscopic
consequences, and which can be avoided in several different ways. It will
not be considered further in the discussion \ here.}. Given the microstates
and dynamics prescribed in this way, the tools of nonequilibrium statistical
mechanics can be applied as well. In particular, the formulation of a
kinetic theory and \ its application to derive the equations of fluid
dynamics for granular systems can be addressed. The objective here is to
summarize the current status of such investigations and to address some of
the early and continuing concerns about the validity of kinetic theory and
hydrodynamics for granular media. Relevant literature prior to 1990 has been
reviewed by Campbell \cite{campbell90rev}. More recent reviews addressing
the statistical mechanical foundations have been given by Ernst \cite%
{ernst00rev} and by van Noije and Ernst \cite{vanNoije01sm}. The reader is
advised to consult the extensive references in these articles. Here, a
personal perspective on the current status of kinetic theory and
hydrodynamics for granular gases is provided as a supplemental orientation
to this rapidly growing field. The references given are selective and
apologies are offered at the outset to the many other important
contributions not recognized \ explicitly.

Contributions to this field come from increasingly diverse areas: chemical
engineering, fluid dynamics, statistical mechanics, condensed matter
physics, computational physics. Each has its priority for problems to be
solved, in some cases focused more on urgent practical concerns while in
others addressing more abstract ''points of principle''. The material
selected for discussion here reflects the author's personal view that a
fundamental (microscopic) basis for rapid flow granular media can be found
within the general framework of nonequilibrium statistical mechanics, and
that its explication will impact practical applications under realistic
conditions. This perspective approaches granular media via idealized models
for which controlled molecular dynamics simulation, many-body theory, and
real experiments reasonably can be performed to answer specific questions
with clear consequences.

\section{Kinetic theory at low density}

It is taken for granted in all of the following that a microscopic
formulation of non-equilibrium statistical mechanics for inelastic, smooth
hard spheres is justified \cite{vanNoije01sm,brey97statmech,dufty00granada}.
Concerns at this level are the same as those for open systems (inherently
non-equilibrium states) with elastic collisions - e.g. the required
''mixing'' property for the dynamics in phase space to justify the relevance
of average values and fluctuations about them. Once it is recognized that
granular states are always ''non-equilibrium'' the distinction between
systems with elastic and inelastic collisions diminishes considerably, as
emphasized below. A first order question is whether the reduction of
non-equilibrium statistical mechanics to a more practical kinetic theory
description can be justified. This is a difficult problem even for the case
of elastic collisions which, as noted above, has been addressed extensively
and successfully for the hard sphere system over the past thirty years. It
is not the purpose here to summarize the status of that research but rather
to explore any {\em additional} questions and concerns for its application
to granular gases arising from the inelasticity of the collisions. The
general conclusion below is that while there are significant differences in
many derived properties of interest, there are few new qualitative
limitations on the equations describing these properties based on analytic
and simulation studies to date. Some of the confusion and disagreement in
the literature is due to differing perceptions of the necessary conditions
for a kinetic theory (e.g., Boltzmann or Enskog kinetic equations), and the
necessary conditions for a hydrodynamic description (e.g., Navier-Stokes or
non-linear rheological equations). In principle and in practice, the
validity of kinetic theory and the validity of hydrodynamics are independent
questions. However, for gases (no compact structures) it is expected that
existence of a kinetic theory should provide the basis for investigating
hydrodynamics as well.

To set the stage for the discussion here, a brief formal ''derivation'' of
the kinetic description for both the distribution function and fluctuations
at low density is given as an illustrative example. A small parameter
expansion of the reduced distribution function leads to a formal solution to
the BBGKY \ hierarchy \cite{dufty95mirim,brey98correl}. The analysis is
similar to an expansion proposed by Grad for the hard sphere gas with
elastic collisions \cite{grad58}. In this approach, there is no reference to
concepts such as ''approach to equilibrium'', Maxwellian distribution, or
''molecular chaos'', and the distinction between inelastic or elastic
collisions plays no explicit role. Thus, superficially at least, it appears
the basis for the low density kinetic theory is the same in both cases.
Closely related results follow from a cluster expansion of the reduced
distribution functions, again with no formal reference to whether the
collisions are elastic or inelastic \cite%
{ernst00rev,vanNoije01sm,vanNoije98ring}.

The s-particle reduced distribution functions, $f^{(s)}(x_{1},\cdots
,x_{s},t)$, obey the BBGKY hierarchy where $x_{i}=\left( {\bf q}_{i},{\bf v}%
_{i}\right) $. A dimensionless form of this hierarchy is obtained by scaling
the space and time with the mean free path $\ell \equiv 1/(n\sigma ^{2})$
and the mean free time $t_{0}\equiv \ell /v_{0}$. Here, $n$ is the density, $%
\sigma $ is the hard sphere diameter, and $v_{0}$ is some characteristic
velocity. Similarly, the reduced distribution functions are scaled with $%
\left( n/v_{0}^{3}\right) ^{s}$. The resulting dimensionless BBGKY hierarchy
has the form 
\begin{eqnarray}
\lefteqn{\left( \partial _{t}+\sum_{i=1}^{s}v_{i}\cdot \nabla _{i}-\lambda
^{2}\sum_{i<j}^{s}\overline{T}(i,j)\right) f^{(s)}(x_{1},\cdots ,x_{s},t)} 
\nonumber \\
&=&\sum_{i=1}^{s}\int dx_{s+1}\,\overline{T}(i,s+1)f^{(s+1)}(x_{1},\cdots
,x_{s+1},t).  \label{2.1}
\end{eqnarray}%
The binary scattering operator $\overline{T}(i,j)$ for a pair of particles $%
\{i,j\}$ is defined by%
\begin{equation}
\overline{T}(i,j)=\int d\hat{{\bf \sigma }}\ \Theta ({\bf g}_{ij}{\bf \cdot }%
\widehat{\text{{\bf $\sigma $}}})({\bf g}_{ij}{\bf \cdot }\widehat{\text{%
{\bf $\sigma $}}})[\alpha ^{-2}\delta ({\bf q}_{ij}-\lambda \widehat{\text{%
{\bf $\sigma $}}})b_{ij}^{-1}-\delta ({\bf q}_{ij}+\lambda \widehat{\text{%
{\bf $\sigma $}}})],  \label{2.2}
\end{equation}%
where $\hat{{\bf \sigma }}$ is a unit vector along ${\bf q}_{ij}={\bf q}_{i}-%
{\bf q}_{j}$ and $d\hat{{\bf \sigma }}$ denotes a two dimensional solid
angle integration over the sphere for particles at contact. Also, ${\bf g}%
_{ij}\equiv {\bf v}_{i}-{\bf v}_{j}$ is the relative velocity, and $%
b_{ij}^{-1}$ is the scattering operator defined for any function $X({\bf v}%
_{i},{\bf v}_{j})$ by 
\begin{equation}
b_{ij}^{-1}X({\bf v}_{i},{\bf v}_{j})\equiv X({\bf v}_{i}^{\prime },{\bf v}%
_{j}^{\prime }).  \label{2.3}
\end{equation}%
The ''restituting'' velocities (i.e., the inverse of those in (\ref{1.1}))
are%
\begin{equation}
{\bf v}_{i}^{\prime }\equiv {\bf v}_{i}-\frac{1+\alpha }{2\alpha }({\bf g}%
_{ij}\cdot \hat{{\bf \sigma }})\hat{{\bf \sigma }},\hspace{0.5cm}{\bf v}%
_{j}^{\prime }\equiv {\bf v}_{j}+\frac{1+\alpha }{2\alpha }({\bf g}%
_{ij}\cdot \hat{{\bf \sigma }})\hat{{\bf \sigma }}.  \label{2.4}
\end{equation}%
The $\alpha $ dependence\ of $\overline{T}(i,j)$ contains all aspects of the
inelasticity, and plays no explicit role in the following expansion.

In this dimensionless form the BBGKY \ hierarchy depends on the single
dimensionless parameter $\lambda \equiv \sigma /\ell =n\sigma ^{3}$, the
ratio of the ''force range'' to the mean free path. This parameter is small
at low density, suggesting an expansion for a solution to the entire
hierarchy as a power series in $\lambda $. The dependence on $\lambda $
occurs explicitly as shown on the left side of (\ref{2.1}) and implicitly
through the finite separation of the colliding particles in $\overline{T}%
(i,j)$. The structural features of the expansion in $\lambda $ are simplest
if it is performed at fixed $\overline{T}(i,j)$. The s-particle reduced
distribution functions are taken to have the representation 
\begin{equation}
f^{(s)}(x_{1},\cdots ,x_{s},t)=f_{0}^{(s)}(x_{1},\cdots ,x_{s},t)+\lambda
^{2}f_{1}^{(s)}(x_{1},\cdots ,x_{s},t)+...  \label{2.5}
\end{equation}%
It is then readily shown that the hierarchy is solved exactly to order $%
\lambda ^{2}$ in the form 
\begin{equation}
f_{0}^{(s)}(x_{1},\cdots ,x_{s},t)=\prod_{i=1}^{s}f_{0}^{(1)}(x_{i},t),
\label{2.6}
\end{equation}%
\begin{equation}
f_{1}^{(s)}(x_{1},\cdots ,x_{s},t)=\sum_{j=1}^{s}\prod_{i\neq
j}^{s}f_{0}^{(1)}(x_{i},t)f_{1}^{(1)}(x_{j},t)+\sum_{i<j}^{s}\prod_{k\neq
i,j}^{s}f_{0}^{(1)}(x_{k},t)G(x_{i},x_{j},t),  \label{2.7}
\end{equation}%
where the expression for $f_{1}^{(s)}$ holds for $s\geq 2$. Thus, the
reduced distribution functions for any number of particles is determined as
a sum of products of the single particle functions $f_{0}^{(1)}(x_{1},t)$
and $f_{1}^{(1)}(x_{1},t)$, and the pair function $G(x_{1},x_{2},t)$. These
are determined from the set of three kinetic equations 
\begin{equation}
\left( \frac{\partial }{\partial t}+{\bf v}_{1}\cdot \nabla _{1}\right)
f_{0}^{(1)}(x_{1},t)=J(x_{1},t\mid f_{0}^{(1)}).  \label{2.8}
\end{equation}%
\begin{equation}
\left( \frac{\partial }{\partial t}+{\bf v}_{1}\cdot \nabla _{1}-I_{1}+{\bf v%
}_{2}\cdot \nabla _{2}-I_{2}\right) G(x_{1},x_{2},t)=\overline{T}%
(1,2)f_{0}^{(1)}(x_{1},t)f_{0}^{(1)}(x_{2},t)  \label{2.9}
\end{equation}%
\begin{equation}
\left( \frac{\partial }{\partial t}+{\bf v}_{1}\cdot \nabla
_{1}-I_{1}\right) f_{1}^{(1)}(x_{1},t)=\int dy_{2}\,\overline{T}%
(1,2)G(x_{1},x_{2},t),  \label{2.10}
\end{equation}%
Here $J(x_{1},t\mid f_{0}^{(1)})$ is the Boltzmann-Bogoliubov collision
operator and $I_{1}$, defined over functions of $x_{1}$, is its linearized
form 
\begin{equation}
J(x_{1},t\mid f_{0}^{(1)})=\int dx_{2}\,\overline{T}%
(1,2)f_{0}^{(1)}(x_{1},t)f_{0}^{(1)}(x_{2},t)  \label{2.10a}
\end{equation}%
\begin{equation}
I_{1}h(x_{1})\equiv \int dx_{2}\,\overline{T}(1,2)\left(
f_{0}^{(1)}(x_{1},t)h(x_{2})+h(x_{1})f_{0}^{(1)}(x_{2},t)\right) .
\label{2.11}
\end{equation}

These low density results (\ref{2.8})-(\ref{2.10}) are remarkably rich. The
leading order distribution function $f_{0}^{(1)}$ is the solution to the
Boltzmann equation. The two particle correlations are generated from the
uncorrelated product of Boltzmann solutions through inelastic binary
collisions $\overline{T}(1,2)f_{0}^{(1)}(x_{1},t)f_{0}^{(1)}(x_{2},t)$ on
the right side of (\ref{2.9}). Finally, corrections to the Boltzmann
solution due to correlations are given by a coupling of the distribution
function to the correlations in (\ref{2.10}) (the so-called ''ring''
recollision effects). In the following sections, it is noted that the
solutions and implications of these kinetic equations can be quite different
for elastic and inelastic collisions. But these differences come from the
equations themselves and should not be interpreted as signatures of their
failure to apply. For example, at $\alpha =1$ a possible solution for an
isolated system is $f_{0}^{(1)}(x_{1},t)\rightarrow f_{M}(v_{1})$, $%
G(x_{1},x_{2},t)=0=f_{1}^{(1)}(x_{1},t)$, where $f_{M}$ is the
Maxwell-Boltzmann distribution. Equation\ (\ref{2.10}) supports $f_{M}$
because energy is conserved, and $G(x_{1},x_{2},t)=0$ because $T(1,2)f_{M}(%
{\bf v}_{1})f_{M}({\bf v}_{2})=0$ for the same reason. Since energy
conservation no longer holds with $\alpha <1$ it is not surprising that the
isolated system does not approach equilibrium, the Maxwellian is not a
stationary solution, and that finite correlations exist. Indeed, the extent
to which such predicted differences agree with observations from molecular
dynamics provide {\em support} for the kinetic theory, not {\em limitations}
on it as is sometimes implied.

Clearly, the above derivation has not restricted this kinetic description to
isolated systems. In fact, the most interesting cases of practical interest
are response to boundary conditions and/or external fields. The similarities
between normal and granular fluids is closest for such ''nonequilibrium''
conditions. Too often, properties of granular gases are contrasted only to
those of the equilibrium state for normal gases. It is important to note
that practical access to the solutions to the above kinetic equations is
possible for a wide range of conditions by direct simulation Monte Carlo
(DSMC) \cite{bird}. Analytic results for complex states have been obtained
using accurate kinetic models for the above equations \cite{brey00model}, as
discussed in Section 7. The benchmark for validity of these kinetic
equations is molecular dynamics (MD) simulation of the idealized granular
fluid microdynamics. Frequently, these simulations are performed at higher
densities than can be supported by the low density approximation. However,
there is a corresponding Enskog approximation that extends the above low
density kinetic theory to moderate densities as well \cite%
{brey97statmech,dufty94}. It is appropriate to emphasize also that the
terminology ''kinetic theory'' is not limited to Boltzmann and Enskog
equations. Indeed, at high densities and long times it is known that there
are important (even dominant) corrections due to correlated many-body
collisions of the type described by (\ref{2.10}) at low density. Thus,
observed deviations from Boltzmann-Enskog \ for $\alpha <1$ do not
necessarily imply a failure of a kinetic theory description in general.

\section{Homogeneous cooling state}

An isolated homogenous fluid with elastic collisions rapidly approaches the
stationary equilibrium Gibbs state after a few collisions. The single
particle distribution function is Maxwellian and there are no correlations
on length scales large compared to the hard sphere diameter. In contrast,
for inelastic collisions no stationary state is possible for the isolated
system. Instead, it is straightforward to show that the temperature (defined
in the usual way from the average kinetic energy) is monotonically
decreasing due to the inelastic collisions. Still, on the same time scale as
the approach to equilibrium for elastic collisions, the granular fluid
approaches a special ''homogeneous cooling state'' (HCS) for which all time
dependence of the distributions occurs only through this cooling
temperature. This implies a scaling form for the single particle distribution%
\begin{equation}
\,f_{0}^{(1)}({\bf v},t)=v_{0}^{-3}(t)n\phi \left( v/v_{0}(t)\right) ,%
\hspace{0.5cm}v_{0}^{2}(t)=2T(t)/m.  \label{3.1}
\end{equation}%
The pair correlations have a similar scaling form. The existence of a
scaling solution and corresponding time dependence of $T(t)$, its approach
from general homogeneous states, the form of $\phi \left( v\right) $, and
the space dependence of correlations in this state provide specific tests
for the kinetic theory. The scaling property implies an exact time
dependence $T(t)=T(0)(1+ct)^{-2}$ (Haff's law) which has been confirmed by
MD \cite{Huthmann00} and by DSMC solution to the Boltzmann and Enskog
equations \cite{brey96hcs}. In both the MD and DSMC the approach to a
scaling solution is observed to occur rapidly after a few collisions,
similar to the approach to equilibrium for normal fluids. The velocity
dependence of $\phi \left( v\right) $ has been calculated approximately from
the the Boltzmann and Enskog equations by an expansion about the Gaussian $%
\left( \alpha =1\text{ form}\right) $ in polynomials \cite%
{goldstein95,vanNoije98hcs}. Deviations from the Gaussian have been measured
by MD and DSMC and compared with a low order truncation of the expansion.
The agreement is excellent for all $0.3<\alpha \leq 1$. Finally, the form of
this scaling distribution has been analyzed from the Boltzmann equation for
large velocities and predicted to have an exponential rather than Gaussian
form \cite{vanNoije98hcs}. This also has been confirmed in detail by DSMC %
\cite{brey99tail} and the effect has been observed by MD simulation as well %
\cite{Huthmann00}.

Recently, the scaling solutions to the Boltzmann equations for a mixture
have been studied \cite{garzo99hcs}. A new effect is the occurrence of
different temperatures for each species, although each with the same cooling
rate given by Haff's law. This difference is a reflection of the failure of
energy equipartition in the HCS, an exact property of the Gibbs state. The
polynomial solutions for the mixture distribution functions can be
constructed in the same way as for the one component fluid. The rapid
approach to this HCS, the dependence of the different temperatures on
species parameters (size, concentration, mass), and the velocity dependence
of the scaling solutions have been confirmed in detail by DSMC \cite%
{montanero01}, for $\alpha \geq 0.5$. To date there have been no MD
simulations for comparison.

Correlations in the HCS are described by Eq.(\ref{2.9}) for $%
G(x_{i},x_{j},t)=f^{(2)}(x_{1},x_{2},t)-f^{(1)}(x_{1},t)f^{(1)}(x_{2},t)$.
The origin of these correlations is the source term in this equation $%
\overline{T}(1,2)\phi \left( v_{1}/v_{0}(t)\right) \phi \left(
v_{2}/v_{0}(t)\right) \neq 0$. This vanishes for elastic collisions because
in that case $\phi $ becomes the Maxwellian, and $\overline{T}(12)$ regains
the detailed balance property. For $\alpha <1$ this finite source gives
non-zero {\em short ranged} spatial correlations $\sim \delta (q_{ij}-\sigma
)$. The generators for the two particle dynamics on the left side of (\ref%
{2.9}) include in their spectrum the inelastic fluid hydrodynamic modes
(discussed below). These modes give rise to {\em long ranged }spatial
correlations in $G(x_{i},x_{j},t)$, including algebraic decay at large
distances for velocity moments corresponding to density, energy, and
momentum correlations. A detailed study of the predictions for these
correlations based on (\ref{2.9}) has been given by Brey et al. \cite%
{brey98correl}. These calculations require analysis of the spectrum for $%
{\bf v}_{1}\cdot \nabla _{1}-I_{1}$ at long wavelengths, and confirm the
existence of hydrodynamic modes as spontaneous fluctuations in the granular
gas. These modes agree with those obtained for the decay of imposed spatial
gradients described in the Section 6, and this spectral analysis can be
viewed as an independent confirmation of hydrodynamics (in effect, this is
Onsager's regression hypothesis extended to granular fluids). The dispersion
relations for these modes as a function of $\alpha $ and wavevector is \
more complex than for normal fluids and the spatial dependence cannot be
obtained from simple scaling without specification of a domain for $\alpha $%
. The calculated space dependence of the correlation functions is in
excellent agreement with DSMC simulation of the \ kinetic equation (\ref{2.9}%
).

Correlations on such macroscopic scales in normal fluids also can be
described by the usual hydrodynamic equations extended to include thermal
fluctuations characterized by white, Gaussian noise. Averaging products of
solutions to these stochastic equations over the noise provides the
correlation functions. van Noije et al. \cite{brito97} have extended this
approach to granular fluids as well. The agreement with the kinetic theory
and DSMC results above is good, except for large inelasticity where the use
of elastic fluid transport coefficients in \cite{brito97} is not justified.
Furthermore, these results are in good agreement with MD simulations. In
summary, the consistency of kinetic theory and hydrodynamics to predict a
new effect for granular gases - long range correlations in an isolated
homogeneous system - appears to be established.

Spatial correlations on short length scales have been described in detail by
Lutsko \cite{lutsko00,lutsko01}. The analysis is based on an exact boundary
condition for hard spheres at contact \cite{lutsko96} 
\begin{eqnarray}
&&\delta \left( {\bf q}_{12}-{\bf \sigma }\right) \left( {\bf \sigma }\cdot 
{\bf g}_{12}\right) \Theta \left( {\bf \sigma }\cdot {\bf g}_{12}\right)
f^{(2)}(x_{1},x_{2},t)d{\bf v}_{1}d{\bf v}_{2}  \nonumber \\
&=&\delta \left( {\bf q}_{12}-{\bf \sigma }\right) \left( -{\bf \sigma }%
\cdot {\bf g}_{12}^{\prime }\right) \Theta \left( -{\bf \sigma }\cdot {\bf g}%
_{12}^{\prime }\right) f^{(2)}(x_{1}^{\prime },x_{2}^{\prime },t)d{\bf v}%
_{1}^{\prime }d{\bf v}_{2}^{\prime }  \label{3.2}
\end{eqnarray}%
The left side describes the flux of particles on the post-collision
hemisphere. The right side is the distribution of particles on the
pre-collision hemisphere as a function of their restituting velocities, and
the equality represents their equality under a two particle collision. Then
using the properties $-{\bf \sigma }\cdot {\bf g}_{12}^{\prime }=\alpha ^{-1}%
{\bf \sigma }\cdot {\bf g}_{12}$ and $d{\bf v}_{1}^{\prime }d{\bf v}%
_{2}^{\prime }=\alpha ^{-1}d{\bf v}_{1}d{\bf v}_{2}$ the half space
distributions at ${\bf q}_{12}={\bf \sigma }$ are seen to be related by 
\begin{equation}
\Theta \left( {\bf \sigma }\cdot {\bf g}_{12}\right)
f^{(2)}(x_{1},x_{2},t)=\alpha ^{-2}b^{-1}\Theta \left( -{\bf \sigma }\cdot 
{\bf g}_{12}\right) f^{(2)}(x_{1},x_{2},t)  \label{3.2a}
\end{equation}%
Lutsko observes that the class of generalized mean spherical approximations
(GMSA) for the equilibrium radial distribution $g\left( {\bf q}_{1}-{\bf q}%
_{2},\alpha =1\right) $ function requires primarily the value of this
function at contact and the equation of state. Consequently the GMSA can be
extended to the HCS for $g\left( {\bf q}_{1}-{\bf q}_{2},\alpha \leq
1\right) $ with appropriate changes in these quantities. The value at
contact is obtained by using the approximation%
\begin{equation}
\Theta \left( -{\bf q}_{12}\cdot {\bf g}_{12}\right)
f^{(2)}(x_{1},x_{2},t)\rightarrow \Theta \left( -{\bf q}_{12}\cdot {\bf g}%
_{12}\right) g\left( {\bf q}_{1}-{\bf q}_{2},1\right)
f^{(1)}(x_{1},t)f^{(2)}(x_{2},t)  \label{3.3}
\end{equation}%
{\em on the pre-collision hemisphere}, i.e. neglect of velocity \
correlations on this hemisphere. The boundary condition (\ref{3.2}) then
gives the remaining half of the distribution on the post-collision
hemisphere {\em with velocity correlations} determined from the binary
collision. The equation of state is determined from the Enskog kinetic
equation which can be obtained in this context as follows \cite{lutsko96}.
The exact first equation of the BBGKY hierarchy (Eq.(\ref{2.1}) with $s=1$),
is written in the equivalent form 
\begin{eqnarray}
\left( \frac{\partial }{\partial t}+{\bf v}_{1}\cdot \nabla _{1}\right)
f_{0}^{(1)}(x_{1},t) &=&\sigma ^{2}\int d\overrightarrow{x}_{2}\int d%
\widehat{\sigma }\;\delta \left( {\bf q}_{12}-\sigma \widehat{\sigma }%
\right) \left( \widehat{\sigma }\cdot {\bf g}_{12}\right) \left[ \alpha ^{-2}%
\widehat{b}_{ij}^{-1}+1\right]  \nonumber \\
&&\times \Theta \left( -\widehat{\sigma }\cdot {\bf g}_{12}\right)
f_{2}(x_{1},x_{2},t).  \label{3.4}
\end{eqnarray}%
This shows that any approximation for $f_{2}(x_{1},x_{2},t)$ need only be
imposed for the pair of particles at contact and on the pre-collision \
hemisphere. In particular, the use of (\ref{3.3}) in (\ref{3.4}) leads to
the Enskog kinetic equation. This route to the Enskog (and hence Boltzmann
equation at low density) provides a precise meaning to the phenomenological
terminology ''molecular chaos assumption '': neglect of two particle
velocity correlations for a pair of particles at contact on the
pre-collision hemisphere. The pressure in the HCS follows from the Enskog
equation by evaluating the trace of the average momentum flux. In this way,
the radial distribution function at contact and pressure are found to be $%
g\left( q_{12}=\sigma ,\alpha \right) =\frac{1+\alpha }{2\alpha }g\left(
q_{12}=\sigma ,\alpha =1\right) $ and $p\left( \alpha \right) =nk_{B}T\left[
1+\frac{\left( 1+\alpha \right) }{3}\pi n\sigma ^{3}g\left( q_{12}=\sigma
,1\right) \right] $, respectively. Using $g\left( q_{12}=\sigma ,\alpha
\right) $ and $p\left( \alpha =1\right) $ in the GMSA Lutsko obtains a model
for $g\left( q_{12},\alpha \right) $ as a function of $q_{12}$, $\alpha $,
and the density. The agreement with MD simulation is very good even at
extreme conditions of high density, $n\sigma ^{3}=0.5$ and strong
dissipation $\alpha =0.5$, for the domain tested $\sigma \leq $ $q_{12}\leq
3\sigma $. This is an indirect confirmation of the molecular chaos
assumption (\ref{3.3}), supporting as well the basis for the Enskog kinetic
equation. In detail, however, there are deviations for $\alpha <0.7$ that
have a system size dependence being larger for the larger \ system (in
particular, the molecular chaos value at contact appears to fail). The
explanation for this is a long wavelength hydrodynamic shear mode becoming
unstable at about $\alpha =0.7$, as discussed in the next section. When this
is compensated for in the simulation the system size dependence no longer
occurs, but significant deviations from the ''molecular chaos''
approximation remain at high density for $\alpha <0.6$. A similar limitation
on the ''molecular chaos'' assumption was observed in a two dimensional MD
simulation \cite{luding98} where the distribution of impact parameters on
the pre-collision hemisphere was found to become non-uniform (particularly
for large impact parameters) at values of the restitution coefficient for
which significant shear fluctuation is observed.

The validity of (\ref{3.3}) has been tested by MD simulation more directly %
\cite{soto00}, although for the limited domain $0.98\leq \alpha \leq 1$, $%
0.05\leq n\sigma ^{3}\leq 0.2$. These authors study a generalization of the
pair correlation function obtained by integrating $f^{(2)}$ over the center
of mass velocity and the magnitude of the relative velocity. This defines a
pair correlation function $g\left( q_{12}=\sigma ,\theta ,\alpha \right) $
depending on the angle between the relative velocity and the relative
coordinate. The MD results for $g\left( q_{12}=\sigma ,\theta ,\alpha
\right) $ as a function $\theta $ for $n\sigma ^{3}=0.1$ and $\alpha =0.98$
show good agreement with the molecular chaos approximation except near $%
\theta =\pi /2$ (large impact parameters). To resolve small differences the
pressure is measured directly from the virial and an order parameter for
pre-collision velocity correlations is measured. The pressure shows a
systematic deviation from the Enskog pressure of less than one percent for
the density range considered. The order parameter for velocity correlations
has a significant system size dependence and an extrapolation of MD results
to large systems is required. Signatures of finite pre-collision velocity
correlations are found in this way, indicating deviations from (\ref{3.3}).
In spite of the good agreement found for $g\left( q_{12}=\sigma ,\theta
,\alpha \right) $, the authors emphasize limitations of the Enskog kinetic
theory based on the pressure and order parameter measurements. To put this
in context, it should be recognized that the same small corrections to the
Enskog equation can occur at these densities also with elastic collisions, 
{\em except} for the equilibrium state. Indeed, they can be calculated from
Eq. (\ref{2.9}) for $\alpha =1$. Once it is recognized that the HCS is a
nonequilibrium state the role of the Enskog kinetic theory as a good
approximation at low and moderate densities is seen to be supported for both
elastic and inelastic collisions. However, it appears from the discussion of
the last paragraph that the range of densities for which this is true
decreases with increased dissipation.

The strongest objections to the above description for properties of the\ HCS
\ and the validity of the Boltzmann equation for granular gases have been
offered in reference \cite{salazar99}. However, both the DSMC simulation and
theoretical analysis presented there have subsequently been criticized as
invalid\cite{elskens99salazar,ernst99salazar}.

Related studies of homogeneous {\em steady} states (HSS) also have been
studied as tests of kinetic theory \cite%
{vanNoije99,bizon99,bizon00,montanero00}. Such steady states are produced by
insertion of a random external force to accelerate the particles and hence
compensate for collisional cooling. This has the advantage of avoiding the
intrinsic time dependence of the HCS, but at the price of introducing
unknown new effects induced by the external force. It should be noted that
the time dependent HCS can be mapped formally onto a stationary state for
time independent study as well, by a suitable change of variables for the
velocities and time \cite{lutsko01}. There are considerably more MD results
for the HSS, with similar conclusions regarding the deviations of the
velocity distribution from Maxwellian, excess population at large
velocities, and velocity correlations that are generally consistent with
kinetic theory as described above for the HCS. Fluctuating Navier-Stokes
order hydrodynamics, including noise from the external random force predicts
well both spatial correlations and fluctuation-renormalization of mean
values \cite{britomode98,vanNoije99}.

This section has focused on the velocity distribution and correlations for
the simplest state of a granular gas, isolated and uniform. Significant
differences are observed from the corresponding equilibrium state for the
same gas with elastic collisions. In some of the literature quoted the
existence of such differences from the equilibrium state interpreted as a
failure of conditions for kinetic theory for granular gases. As the
discussion above shows, it is quite the other way around - these differences
are quite well predicted from the kinetic theory. Deviations from Enskog
kinetic theory at higher densities are expected due to correlated
collisions, just as for fluids with elastic collisions \cite{vesely90}.
There is some indication that the effects of correlated collisions are
enhanced at strong dissipation, and hence appear at lower densities, due to
unstable fluctuations in the transverse flow field as discussed in the next
section. It would be interesting to translate this qualitative statement
into a more quantitative one by careful comparison of such effects for
normal fluids in a nonequilibrium state with those of the HCS, since they
are certainly present in both cases.

\section{Instabilities and clustering}

It was first observed in MD simulations that the HCS is unstable to
sufficiently long wavelength perturbations \cite{goldhirsh93}. For systems
large enough to support such spontaneous fluctuations the HCS becomes
inhomogeneous at long times. In MD simulations the inhomogeneities may grow
by the formation of clusters, ultimately aggregating to a single large
cluster \cite{luding99}; if cluster growth is suppressed a vortex field may
grow to the system size where periodic boundary conditions can induce a
transition to a state of macroscopic shear. The mechanism responsible for
the growth of \ inhomogeneities can be understood at the level of
Navier-Stokes hydrodynamics where linear stability analysis shows two shear
modes and a ''heat'' mode to be unstable. Qualitatively \cite{goldhirsh93},
spontaneous vortex fluctuations of sufficient size grow due to this
instability and provide excess local temperature and pressure; a particle
flux is then established from high to low shear domains increasing the
density and cooling rate; this further enhances the pressure difference to
continue the particle flow. A more quantitative description poses a
challenging test for both kinetic theory and hydrodynamics.

Perhaps the simplest signature of inhomogeneities is a deviation of the
global temperature from the \ form given by Haff's law. In the presence of
local flow fields the average kinetic energy for the system decays more
slowly than that implied by the homogenous scaling form (\ref{3.1}). If the
local flow fields are assumed to be dominated by vortex flow, the decay law
can be calculated from hydrodynamics \cite{britomode98}. The agreement of
the slower decay predicted in this way with MD simulation is excellent,
confirming both the instability and the relevance of hydrodynamics for its
description. Another signature of vortex growth is provided by the time
dependence of the transverse velocity field spatial correlations. As
discussed above they can be calculated from the kinetic equation (\ref{2.9})
as was done in references \cite{brey98correl,brey98correl2}, or from
fluctuating hydrodynamics as in references \cite{brito97,britomode98}. The
amplitude of the correlations in the transverse velocity field is seen to
increase by an order of magnitude on the time scale for growth of the
hydrodynamic shear mode. The agreement between MD, DSMC, kinetic theory, and
hydrodynamics is very good.

A more complete investigation of the mechanism for cluster formation has
been carried out by Brey and collaborators \cite{brey99cluster}, based on
DSMC solution to the Boltzmann equation and an approximate solution to the
nonlinear Navier-Stokes hydrodynamic equations derived from it (Section 6).
The linear stability analysis for the hydrodynamic equations provides a
prediction for the critical wavelength for the instability as a function of $%
\alpha $. This was measured by DSMC for the energy at fixed $\alpha $ to
determine the smallest system size for which secular growth occurs. The
agreement between hydrodynamics and simulation for this critical wavelength
is excellent for all $0.7\leq \alpha \leq 0.9$, i.e. including strong
dissipation. Next, the evolution of an initial flow field with sinusoidal
space dependence of wavelength $\lambda $ was considered. As expected from
the linear stability analysis, the transverse component grows in time
preserving the same imposed wavelength. In contrast the initially uniform
density and energy becomes sinusoidal at {\em half }the original wavelength,
suggesting a bilinear coupling of these fields with the transverse flow
field. Indeed, this is consistent with the mechanism proposed by Goldhirsch,
Tan, and Zanetti \cite{goldhirsh93} described above, where the hydrodynamic
viscous heating provides such a bilinear coupling in the temperature
equation. A detailed comparison between a nonlinear hydrodynamic solution
and DSMC for the growth of this second harmonic in the density and
temperature shows excellent agreement \cite{brey99cluster}. This confirms
that the mechanism for initial cluster growth is initiated by the unstable
shear mode and \ its nonlinear coupling to the density.

There are also MD confirmations of a hydrodynamic description for the
instabilities. Deltour and Barrat \cite{deltour97} study the onset of shear
and cluster instabilities for a two dimensional system, showing good
agreement with predictions of linear hydrodynamics. A nonlinear \
hydrodynamic analysis has been compared with MD simulation at much weaker
dissipation, $\alpha \geq 0.95$ \cite{sotoshear00}. The density and
temperature are assumed to be slaved by the linearly unstable transverse
velocity field and quadratic nonlinearities are retained. For fixed system
size there is a critical value of $\alpha $ below which the nonlinear
equations support a stable inhomogeneous macroscopic shearing state. The
critical value of $\alpha $ and the values of the Fourier components for the
flow field, density, and temperature are found to be in very good agreement
with the MD simulation results.

In summary, for sufficiently large systems the HCS develops spatial
inhomogeneities at times long compared to that required for growth of a
local shear fluctuation. The existence, onset, and early evolution of the
instability is well-described by Navier-Stokes order hydrodynamics. The
kinetic equation also can be used to explore the late stage evolution as
well via DSMC. It appears \cite{brey99cluster} that the \ density field
approaches an inhomogeneous steady form, independent of initial conditions,
for fixed $\alpha $. This suggests that the isolated system supports a more
complex, but stable, state whose form is yet to be clarified. MD simulations
do not yet confirm this, perhaps due to the smaller system sizes and time
scales considered.

\section{Hydrodynamics}

Consider now a spatially inhomogeneous state, created either by initial
preparation or by boundary conditions. In the bulk, there are local balance
equations for the density $n({\bf r},t)$, temperature $T({\bf r},t)$ (or
energy density), and flow velocity ${\bf U}({\bf r},t)$%
\begin{equation}
D_{t}n+n\nabla \cdot {\bf U}=0,  \label{5.1}
\end{equation}%
\begin{equation}
D_{t}T+\frac{2}{3nk_{B}}\left( P_{ij}\partial _{j}U_{i}+\nabla \cdot {\bf q}%
\right) =-T\zeta ,  \label{5.2}
\end{equation}%
\begin{equation}
D_{t}U_{i}+(mn)^{-1}\partial _{j}P_{ij}=0,  \label{5.3}
\end{equation}%
where $D_{t}=\partial _{t}+{\bf U}\cdot \nabla $ is the material derivative, 
$P_{ij}({\bf r},t)$ is the pressure tensor and ${\bf q}({\bf r},t)$ is the
heat flux. The form of these balance equations is the same as for fluids
with elastic collisions except for the source term on the on the right side
of (\ref{5.2}) due to the dissipative collisions, where $\zeta \propto
\left( 1-\alpha ^{2}\right) $ is identified as the cooling rate. \ Haff's
law follow's directly from (\ref{5.2}) for spatially homogeneous states and
the scaling law $\zeta \propto \sqrt{T}$. The fluxes $P_{ij}$, ${\bf q}$ \
and the cooling rate $\zeta $ are given as explicit low degree moments of
the distribution functions $f^{(1)}(x_{1},t)$ and $f^{(2)}(x_{1},x_{2},t)$.

These balance equations are an exact consequence of the Liouville dynamics
for the system. Their utility is limited without further specification of $%
P_{ij}$, ${\bf q}$, and $\zeta $ which, in general, have a complex space and
time \ dependence. However, for a fluid with elastic collisions this
dependence ''simplifies'' on sufficiently large space and time scales where
it is given entirely through a functional dependence on the fields $n$, $T$,
and ${\bf U}$. The resulting functional dependencies of $P_{ij}$ and ${\bf q}
$ on these fields are called constitutive equations and their discovery can
be a difficult many-body problem. The above balance equations, together with
the constitutive equations, become \ a closed set of equations for $n$, $T$,
and ${\bf U}$ are called {\em hydrodynamic equations}. This is the most
general and abstract notion of hydrodynamics, which encompasses both the
Navier-Stokes form for small spatial gradients and more general forms for
nonlinear rheological transport. The primary feature of a hydrodynamic
description is the reduction of the description from many microscopic
degrees of freedom to a set of equations for only five local fields.

How does this reduction come about? For the case of elastic collisions the
chosen fields are local densities of globally conserved number, energy, and
momentum. Consequently, the dynamics of these variables can be made as slow
as desired by considering long wavelength phenomena. In this way a time
scale can be chosen such that all other excitations have decayed to zero,
leaving a time dependence for all properties only through these conserved
densities. In practice, the hydrodynamic fields dominate for times large
compared to the mean free time, and for wavelengths large compared to the
mean free time. This explains the wide domain of applicability for a
hydrodynamic description of physical phenomena.

The extension of these \ ideas to granular flow raises several questions: 1)
since energy is not conserved, why should the equation for $T$ be included
in the set of slow hydrodynamic fields?, 2) is the new time scale associated
\ with the cooling rate $\zeta $ microscopic or macroscopic?, 3) without any
equilibrium Gibbs state, what is the reference state toward which these
fields are relaxing? These issues are quite different from the those
regarding the validity of a kinetic theory, so to address them consider
values of the density and restitution coefficient for which the Enskog
kinetic equation is reliable. This is given by the exact first BBGKY
hierarchy equation (\ref{3.4}) with the molecular chaos assumption \cite%
{brey97statmech} 
\begin{equation}
\left( \frac{\partial }{\partial t}+{\bf v}_{1}\cdot \bbox {\nabla}%
_{1}\right) f({\bf r}_{1},{\bf v}_{1},t,)=J_{E}[{\bf r}_{1},{\bf v}%
_{1}|f(t)],  \label{5.4}
\end{equation}%
where $J_{E}$ is the Enskog collision operator, 
\begin{eqnarray}
J_{E}[{\bf r}_{1},{\bf v}_{1}|f(t)] &\equiv &\sigma ^{2}\int d{\bf v}%
_{2}\int d\widehat{\bbox {\sigma }}\,\Theta (\widehat{\bbox {\sigma }}\cdot 
{\bf g})(\widehat{\bbox {\sigma }}\cdot {\bf g})\left\{ \alpha ^{-2}f^{(2)}(%
{\bf r}_{1},{\bf r}_{1}-\bbox {\sigma},{\bf v}_{1}^{\prime },{\bf v}%
_{2}^{\prime },t)\right.  \nonumber \\
&&\left. -f^{(2)}({\bf r}_{1},{\bf r}_{1}+\bbox{\sigma},{\bf v}_{1},{\bf v}%
_{2},t)\right\} ,  \label{5.5}
\end{eqnarray}%
\begin{equation}
f^{(2)}({\bf r}_{1},{\bf r}_{2},{\bf v}_{1},{\bf v}_{2},t)\equiv g({\bf r}%
_{1},{\bf r}_{2}|n(t))f({\bf r}_{1},{\bf v}_{1},t,)f({\bf r}_{2},{\bf v}%
_{2},t).  \label{5.6}
\end{equation}%
(Here and in the following the superscript on $f^{(1)}$ will be suppressed
to simplify the notation). The balance equations in the form (\ref{5.1}) - (%
\ref{5.3}) above follow directly from this equation without further
approximation. Similarly, $P_{ij}$, ${\bf q}$, and $\zeta $ are found to be
bilinear functionals of $f$ \cite{brey97statmech} 
\begin{eqnarray}
P_{ij} &=&\int d{\bf v}\,mV_{i}V_{j}\,f({\bf r},{\bf v},t)+\frac{1+\alpha }{4%
}m\sigma ^{3}\int \,d{\bf v}_{1}\int \,d{\bf v}_{2}\int d\Omega \ \Theta (%
\widehat{\bbox{\sigma }}\cdot {\bf g})((\widehat{\bbox{\sigma}}\cdot {\bf g}%
)^{2}\widehat{\sigma }_{i}\widehat{\sigma }_{j}  \nonumber \\
&&\times \ \int_{0}^{1}d\lambda f^{(2)}({\bf r}-(1-\lambda )\bbox{ \sigma},%
{\bf r}+\lambda \bbox{\sigma},{\bf v}_{1},{\bf v}_{2},t),  \label{5.7}
\end{eqnarray}%
\begin{eqnarray}
{\bf q} &=&\int d{\bf v}\,\frac{1}{2}mV^{2}{\bf V}f({\bf r},{\bf v},t)+\frac{%
1+\alpha }{4}m\sigma ^{3}\int \,d{\bf v}_{1}\int \,d{\bf v}_{2}\int d\Omega
\ \Theta (\widehat{\bbox{\sigma }}\cdot {\bf g})((\widehat{\bbox{\sigma }}%
\cdot {\bf g})^{2}({\bf G}\cdot \widehat{\bbox {\sigma }})\widehat{%
\bbox{\sigma }}  \nonumber \\
&&\times \ \int_{0}^{1}d\lambda f^{(2)}({\bf r}-(1-\lambda )\bbox{ \sigma},%
{\bf r}+\lambda \bbox{\sigma},{\bf v}_{1},{\bf v}_{2},t),  \label{5.8}
\end{eqnarray}%
\begin{equation}
\zeta =\left( 1-\alpha ^{2}\right) \frac{\beta m\sigma ^{2}}{12nT}\int \,d%
{\bf v}_{1}\,\int \,d{\bf v}_{2}\,\int d\Omega \,\Theta (\widehat{\bbox{
\sigma }}\cdot {\bf g})(\widehat{\bbox{\sigma }}\cdot {\bf g})^{3}\ f^{(2)}(%
{\bf r},{\bf r}+\bbox{\sigma},{\bf v}_{1},{\bf v}_{2},t),  \label{5.9}
\end{equation}%
where ${\bf V}={\bf v}-{\bf U}({\bf r},t)$ is the peculiar velocity and $%
{\bf G}=\frac{1}{2}({\bf V}_{1}+{\bf V}_{2})$. Equations (\ref{5.1}) - (\ref%
{5.9}) provide an unambiguous basis to study the existence and criteria for
a hydrodynamic description. Although these equations are complicated, DSMC
provides direct numerical access to solutions to the kinetic equation to
determine the fields and fluxes without any assumptions regarding
hydrodynamics for critical tests and benchmarks.

Hydrodynamics results from the balance equations supplemented with
constitutive equations for $P_{ij}$, ${\bf q}$, and $\zeta $. Since the
former are exact, the critical issue is the existence and form of the
constitutive equations. It is clear from (\ref{5.7}) - (\ref{5.9}) that they
will be obtained if the Enskog equation admits a ''normal'' solution, whose
space and time dependence occurs entirely through its functional dependence
on the fields 
\begin{equation}
f({\bf r},{\bf v},t)=F\left( {\bf v\mid }n,T,{\bf U}\right)  \label{5.10}
\end{equation}%
The fluxes and cooling rate then inherit this space and time dependence and
become constitutive equations. The space and time dependence of the fields
follows from solution to the resulting hydrodynamic equations to complete
the self-consistent description of $F$. A hydrodynamic description for
granular gases can be justified if it can be shown that such a ''normal''
solution exists, and that it represents a wide class of more general
solutions on a large space and time scale. This is a difficult task even for
the case of elastic collisions, and detailed results exist only for a few
special nonequilibrium states using model kinetic equations (see Section 7
below). For the Boltzmann and Enskog equations partial answers exist for
gases with elastic collisions and states with small spatial gradients. In
that case, the Chapman-Enskog method \cite{ferziger72} generates the normal
solution explicitly by a systematic expansion in the small gradients . To
lowest order the resulting constitutive equation for the pressure tensor is
Newton's viscosity law, while that for the heat flux is Fourier's law. The
pressure and associated transport coefficients in these expressions (shear
and bulk viscosities, thermal conductivity) also are given by the method in
\ terms of the Enskog collision operator. The balance equations become the
familiar Navier-Stokes hydrodynamic equations. The dominance of the
hydrodynamic description at large space and time scales is justified by a
study of the spectrum of the linearized Enskog equation. It can be shown
that there are five smallest eigenvalues (hydrodynamic modes) that scale
with the wavelength such that they are isolated from all other eigenvalues %
\cite{mclennan89} for wavelengths long compared to the mean free path. Thus
a general excitation of the system will be dominated by the hydrodynamic
modes after a short transient time during which all other excitations decay.
It is verified that the spectral hydrodynamic modes are the same as those
obtained from the linearized Navier-Stokes equations obtained by the
Chapman-Enskog method, at long wavelengths.

While questions of convergence of the Chapman-Enskog solution and the nature
of the constitutive equations and boundary \ conditions beyond Navier-Stokes
order remain open, in general, the conceptual basis for hydrodynamics in
fluids with elastic collisions seems clear and convincing. In the next
section, the extent to which a similar analysis applies for granular gases
is discussed.

\section{Local HCS and Navier-Stokes equations}

The mean free path $\ell =1/n\sigma ^{2}$ is a characteristic microscopic
length both for fluids with elastic and inelastic collisions. Consider a
state for which the spatial variations of $n$, $T$, and ${\bf U}$ are small
on the scale of the mean free path (e.g., $\ell \nabla \ln n<<1$). Then it
is expected that the functional dependence of the normal solution on the
hydrodynamic fields can be made local in space\ through a Taylor series
expansion about the point ${\bf r}$ and time $t$ for which the distribution
function is being evaluated. Similarly, since the solution is normal, all
time derivatives occur only through the fields which obey the balance
equations (\ref{5.1})--(\ref{5.3}). For fluids with elastic collisions,
these time derivatives are proportional to gradients of the fluxes which in
turn are proportional to the small gradients of the fields. As discussed
above, this is the basis for the dominance of hydrodynamics at long times
and long wavelengths. The first difference for granular fluids appears at
this point in the analysis since the time derivative of the temperature is
not simply proportional to the gradients but also has a contribution from
the cooling rate $-\zeta $. Since this is proportional to $\left( 1-\alpha
^{2}\right) $ it can be made small for sufficiently weak dissipation,
although this is {\em not a condition} for the Chapman-Enskog expansion
described below. It is only necessary that $\zeta $ remains smaller than the
decay rates for the non-hydrodynamic excitations at large dissipation, a
condition for the dominance of a closed set of equations including the
temperature at late times.

Let $\epsilon $ denote a formal small ''uniformity'' parameter measuring the
small spatial gradients in the fields (e.g. a term of order $\epsilon $ is
of first order in a hydrodynamic gradient, $\epsilon ^{2}$ is either a
product of two first degree hydrodynamic gradients or one second degree
hydrodynamic gradient). As anticipated above, there is no restriction that
the cooling rate $\zeta $ be of order $\epsilon $; only the spatial
gradients are being ordered by the uniformity parameter. The distribution
function, collision operator, and time derivative are given by the
representations 
\begin{equation}
F=F^{(0)}+\epsilon F^{(1)}+\cdots ,\hspace{0.3in}J_{E}=J^{(0)}+\epsilon
J^{(1)}+\cdots ,\hspace{0.3in}\partial _{t}=\partial _{t}^{(0)}+\epsilon
\partial _{t}^{(1)}+\cdots  \label{6.1}
\end{equation}%
The coefficients in the time derivative expansion are identified from the
balance equations with a similar expansion for $P_{ij}$, ${\bf q}$, and $%
\zeta $ generated through their definitions (\ref{5.7}) - (\ref{5.9}) as
functionals of $F$. The leading term $F^{(0)}$ is further constrained to
have the same moments with respect to $1,v^{2},$ and ${\bf v}$ as the full
distribution $F$. To zeroth order in $\epsilon $ the macroscopic balance
equations become $\partial _{t}^{(0)}n=0,\quad \partial _{t}^{(0)}{\bf u}%
=0,\quad T^{-1}\partial _{t}^{(0)}T=-\zeta ^{(0)}$, and the time derivative
in the Enskog kinetic equation to this order can be evaluated as 
\begin{equation}
\partial _{t}^{(0)}F^{(0)}=-\zeta ^{(0)}T\partial _{T}F^{(0)}=\frac{1}{2}%
\zeta ^{(0)}\bbox {\nabla}_{V}{\bf \cdot }\left( {\bf V}f^{(0)}\right)
\label{6.2}
\end{equation}%
The first equality follows from the normal form of $F^{(0)}$, and the second
equality follows from dimensional analysis which requires the form 
\begin{equation}
F^{(0)}=n({\bf r},t)v_{0}^{-3/2}\phi (V/v_{0}),\hspace{0.3in}v_{0}^{2}=2T(%
{\bf r},t)/m  \label{6.3}
\end{equation}%
The dependence on the magnitude of ${\bf V}={\bf v}-{\bf U(r,}t{\bf )}$ is
due to the requirement that to zeroth order in $\epsilon $ the distribution
function must be isotropic with respect to the peculiar velocity.

The Enskog kinetic equation to zeroth order in $\epsilon $ determines the
velocity dependence of $F^{(0)}$ 
\begin{equation}
\frac{1}{2}\zeta ^{(0)}\bbox {\nabla}_{V}{\bf \cdot }\left( {\bf V}%
F^{(0)}\right) =J^{(0)}[F^{(0)},F^{(0)}].  \label{6.4}
\end{equation}%
Comparison of (\ref{6.3}) with (\ref{3.1}) shows a close similarity with the
HCS. In fact, Eq. (\ref{6.4}) is the same as that for the HCS so the scaling
function $\phi (x)$ is exactly the same\footnote{%
In Section 3 the Boltzmann equation is considered, while here the Enskog
equation is used. However, for the HCS and to zeroth order in $\epsilon $
the Enskog and Boltzmann collision operators differ only by a constant which
can be absorbed in $\zeta ^{(0)}$.}. Here, however, $F^{(0)}$ is normal so
the density, temperature, and flow velocity of the HCS must be replaced by
their corresponding exact values for the {\em inhomogeneous} state
considered, as indicated explicitly in (\ref{6.3}). This is referred to as
the {\em local }HCS distribution (a misnomer, since it is no longer
homogeneous). The fluxes can be calculated to this order using only the
symmetry of this distribution with the results ${\bf q}^{(0)}=0$, $%
P_{ij}=p\delta _{ij}$. The pressure $p$ is the same as that for the HCS, $%
p=nk_{B}T\left[ 1+\frac{1+\alpha }{3}\pi n\sigma ^{3}g\left( q_{12}=\sigma
\right) \right] $, except now evaluated at the true nonuniform temperature
and density. The evaluation of the cooling rate $\zeta ^{(0)}$ requires the
detailed form for the local HCS, which can be obtained approximately by the
\ polynomial expansion method described in Section 3. At low density, i.e.
the Boltzmann limit, the resulting hydrodynamic equations are the
counterpart of the Euler equations for a fluid with elastic collisions:
first order partial differential equations in space and time. The only
qualitative difference is the source term $-T\zeta ^{(0)}$ in the
temperature equation. At finite density, the Enskog equation gives an
additional contribution to the cooling rate proportional to ${\bf \nabla
\cdot U}$ at the next order in the Chapman-Enskog method, which must be
retained as well at Euler order (hydrodynamic equations to first order in
the spatial gradients). The dense fluid Euler equations were first obtained
from a consistent application of the Chapman-Enskog method by Goldstein and
Shapiro \cite{goldstein95}.

At this point the most significant differences in the application of the
Chapman-Enskog method to granular gases have been exposed and it is an
appropriate point to address some misconceptions. Early derivations of
hydrodynamics from kinetic theory used the local Maxwellian for the
reference state $F^{(0)}$, as is done in the case of elastic collisions \cite%
{jenkins83,lun84,jenkins85}. Since the Maxwellian is only a good
approximation for granular fluids at weak dissipation the applicability of
the resulting hydrodynamics also has this limitation to weak dissipation.
Similarly, it has been argued that hydrodynamics is a process of relaxation
towards equilibrium and therefore can apply only if the granular gas is
close to the local Maxwellian, implying the same limitation. However, the
local Maxwellian is not a representation of the equilibrium state but rather
of the macroscopic non-equilibrium state through its dependence on the
hydrodynamic fields. Only its velocity dependence is equilibrium-like.
Furthermore, the analysis here shows that even this velocity dependence is
not free to be chosen, neither for computational convenience nor to match
conceptual bias from experience for fluids with elastic collisions. Rather
it is determined by the expansion itself. For elastic collisions the left
side of (\ref{6.4}) vanishes and the solution is indeed the {\em local} \
Maxwellian as a consequence of detailed balance. The physical interpretation
is that a wide class of solutions for spatially inhomogeneous states evolve
in two stages. During a short transient period of the order of the mean free
time (the kinetic stage), the velocity distribution approaches closely the
local Maxwellian and becomes normal as in (\ref{6.3}). Subsequently, on a
longer time scale the space and time dependence of the distribution occurs
only through the fields that are governed by hydrodynamic equations. A
similar interpretation holds for the granular gas. During the kinetic stage
the velocity distribution rapidly approaches a normal solution close to the
local HCS, now determined from (\ref{6.4}) with the non-zero contribution
from the cooling rate. The velocity distribution is no longer Maxwellian and
the balance equations do not conserve energy, but the conceptual basis for
this two stage relaxation and the mathematical implementation of the
Chapman-Enskog method for the normal solution is not compromised by these
changes.

It also has been argued that the instabilities of the HCS limit the time
scale for validity of hydrodynamics. This is a valid question for
calculating the linear response of the actual HCS to small spatial
perturbations. In that case a solution to the kinetic equation is sought by
expanding about an unstable state. While the derived linear hydrodynamic
modes may apply through the early stages of instability, eventually the
growing perturbation will invalidate the description. The Chapman-Enskog
method generates a quite different solution, which agrees with linear
response only when the local HCS is linearized about the HCS. The method
generates \ the velocity dependence of a spatially {\em inhomogeneous }%
distribution from the Enskog equation through its parameterization by the
hydrodynamic fields. Its space and time dependence is governed through these
fields by the full nonlinear hydrodynamic equations. Consequently, it is
capable of describing late stage stabilization of linear instabilities as
long as spatial gradients remain relatively small. Examples have been given
in the previous section for linear instabilities associated with vortex
fluctuations that appear to evolve nonlinearly to stationary inhomogeneous
states. Similar situations occur for fluids with elastic collisions, where
the hydrodynamic equations describe well a wide range of bifurcation
phenomena. In such cases both Euler level nonlinearities and the
nonequilibrium state dependence of the transport coefficients (temperature
and density) can be important, and both are incorporated in the
Chapman-Enskog method.

Mathematically, the changes in this method for granular gases arise from the
fact that the time derivative of the temperature does not vanish to lowest
order in $\epsilon $, as it does for a gas with elastic collisions. In fact,
the reference state $F^{(0)}$ incorporates the time dependence of the
temperature even for strong dissipation. It has been remarked that the
validity of the Chapman-Enskog expansion is questionable for such a time
dependent reference state \cite{goldhirsh01}. However, since $F^{(0)}$ is
normal, it necessarily has the time dependence of all hydrodynamic fields
even in the case of elastic collisions. The primary difference for granular
gases is the introduction of a new time scale $1/\zeta ^{(0)}$ in the
reference state. However, there is nothing {\em a priori} inconsistent with
a description of slow spatial decay towards a time dependent reference
state. \ Instead, the key question is whether the new time scale is smaller
than that of other excitations active during the kinetic stage and
negligible on the longer time scale. If so, the inclusion of the temperature
as one of the hydrodynamic fields is justified for these long times. Further
comment on this is given in Section 8.

Implementation of the Chapman-Enskog method to the first order in $\epsilon $
is now straightforward and has been carried out in detail and without
approximation recently for the Boltzmann equation \cite{brey98boltz} and for
the Enskog equation \cite{garzo00enskog}; the case of a two component
mixture is considered in \cite{garzo01mix}. The constitutive equations for
the one component fluid found to this order are%
\begin{equation}
P_{ij}\rightarrow p\delta _{ij}-\eta \left( \partial _{j}U_{i}+\partial
_{i}U_{j}-\frac{2}{3}\delta _{ij}\nabla \cdot {\bf U}\right) -\gamma \delta
_{ij}\nabla \cdot {\bf U},\hspace{0.3in}  \label{6.5}
\end{equation}%
\begin{equation}
{\bf q}\rightarrow -\kappa \nabla T-\mu \nabla n,  \label{6.6}
\end{equation}%
\begin{equation}
\zeta ^{(1)}=\frac{p-p^{k}}{p^{k}}\psi \nabla \cdot {\bf U.}  \label{6.7}
\end{equation}%
The form of the pressure tensor is the same as that for fluids with elastic
collisions, where $\eta \left( n,T,\alpha \right) $ is the shear viscosity
and $\gamma \left( n,T,\alpha \right) $ is the bulk viscosity, depending on
the restitution coefficient as well as the local density and temperature.
The heat flux is similar to Fourier's law, where $\kappa \left( n,T,\alpha
\right) $ is the thermal conductivity. However, there is a new transport
coefficient $\mu \left( n,T,\alpha \right) $ coupling the heat flux to a
density gradient, and a coefficient $\psi \left( n,T,\alpha \right) $
coupling the cooling rate to the divergence of the velocity field. These
latter two coefficients vanish at $\alpha =1$. \ The transport coefficients
are given in terms of solutions to inhomogeneous integral equations
involving the linearized Enskog operator. Solubility conditions for the
existence of solutions have been proven and approximate solutions in terms
of polynomial expansions have been obtained for practical purposes, just as
for the case of elastic collisions \cite{garzo00enskog}. Consequently the
transport coefficients are known as explicit functions of the density,
temperature, and restitution coefficient. The Chapman-Enskog method places
no explicit restriction on these parameters, so their domain of
applicability with respect to density and dissipation is the same as that
for the underlying Enskog kinetic equation. In particular, there is no
restriction to weak dissipation introduced by the method.

The hydrodynamic equations associated with (\ref{6.5}) - (\ref{6.7}) are the
Navier-Stokes equations or Newtonian hydrodynamics (constitutive equations
that are linear in the gradients). Under some circumstances large gradients
occur and more complex constitutive equations are required (e.g., shear flow
discussed below). The need for more complex constitutive equations does not
signal a breakdown of hydrodynamics \cite{tan98}, only a failure of the
Navier-Stokes approximation \cite{dufty99comment}. Although the
Chapman-Enskog method can be carried out to second order in $\epsilon $
(Burnett order), it is likely that failure of the Navier-Stokes
approximation signals the need for other methods to construct the normal
solution that are not based on a small gradient expansion. There are reasons
to expect that such non Newtonian transport may be more prevalent for
granular flows \cite{goldhirsh01}.

The validity of Navier-Stokes hydrodynamics and the dependence of the
transport coefficients on the restitution coefficient has been verified in a
number of simulations, both DSMC and MD, giving good agreement with the
predictions from the Chapman-Enskog method. The tests at low density based
on the Boltzmann equation have been review recently by Brey and Cubero \cite%
{brey01rev}. The details described there will not be repeated; only a
summary of the properties studied and the results obtained is given. The
Navier-Stokes equations described above, with the low density transport
coefficients as explicit functions of $\alpha $ are considered in reference %
\cite{brey99hydro}. Using DSMC the evolution of an initial sinusoidal
transverse component of the velocity field is measured. After a few
collisions it is found to decay according to shear diffusion, characterized
by the shear viscosity. In this way the shear viscosity is measured as a
function of $\alpha $ for $\alpha >0.6$, showing very good agreement with
the results from the Chapman-Enskog method. Similarly, the longitudinal
modes were excited by an initial sinusoidal perturbation of the density. The
evolution of the density and temperature fields, changing an order of
magnitude over $30$ collision times, is in excellent agreement with the
theory using transport coefficients evaluated at the chosen $\alpha =0.7$
(no adjustable parameters). Finally a similar perturbation to excite the
diffusive mode of a labelled particle allowed measurement of the
self-diffusion coefficient in two dimensions; a second measurement of the
mean square displacement was made for comparison as well \cite{brey00diffus}%
. Both results are in excellent agreement with the Chapman-Enskog prediction
for $\alpha >0.6$ \ and with MD results performed for $\alpha >0.7$. More
recently, the self-diffusion coefficient has been compared with MD
simulation at finite densities \cite{lutskoD01}. The agreement is very good
for $\alpha >0.6$ and $n\sigma ^{3}\leq 0.25$. Significant deviations are
observed at higher densities and smaller $\alpha $. The mean square
displacement remains linear in time so that hydrodynamic diffusion still
occurs, although its description by the Enskog equation is no longer
accurate. Presumably, this is due to the enhanced shear fluctuations at
larger densities and smaller $\alpha $ discussed in Section 4. The
dependence of the shear viscosity on density and $\alpha $ has been compared
to results from a DSMC measurement of viscous heating \cite%
{montanero01viscosity}. The agreement with the Chapman-Enskog prediction is
excellent for $\alpha >0.6$ and $n\sigma ^{3}\leq 0.5$ (in this case the
shear fluctuations are implicitly suppressed by the method). These results
show clearly the direct verification of hydrodynamics and the quantitative
predictions for transport coefficients from kinetic theory for states with
small spatial gradients but including strong dissipation.

\section{Kinetic models and exact results}

In the last Section a precise question was posed: does the Enskog kinetic
equation for a granular gas imply a corresponding hydrodynamic description
on appropriate space and time scales. For Newtonian flows and relatively
simple nonequilibrium states, the answer appears to be ''yes''. It is
possible and of great interest to explore non Newtonian flows as well (e.g.
steady shear flow), where the Navier-Stokes equations no longer apply. While
this is possible using DSMC, the Chapman-Enskog method is no longer
appropriate to construct the normal distribution and associated constitutive
equations. This problem has been addressed for the Boltzmann equation with
elastic collisions by considering model kinetic equations with structurally
simpler collision operators that allow more penetrating analysis. In this
way a small set of exact normal solutions has been obtained for spatially
inhomogeneous states far from equilibrium, providing results in
semi-quantitative agreement with DSMC studies of the underlying Boltzmann
equation \cite{duftymodel}. Recently, the method of kinetic models has been
proposed for granular gases based on the Boltzmann and Enskog kinetic
equations \cite{brey99model3}. In the former case the Boltzmann equation is
rewritten as 
\begin{equation}
\left( \partial _{t}+{\bf v\cdot \nabla }\right) f-\frac{\zeta }{2}\frac{%
\partial }{\partial {\bf v}}\cdot \left( {\bf V}f\right) =J^{^{\prime }}(f),
\label{7.1}
\end{equation}%
\begin{equation}
J^{^{\prime }}(f)\equiv J(f)-\frac{\zeta }{2}\frac{\partial }{\partial {\bf v%
}}\cdot \left( {\bf V}f\right) .  \label{7.2}
\end{equation}%
where $J(f)$ is the Boltzmann collision operator defined in (\ref{2.10a}).
The velocity derivative term on the left side compensates for the time
dependence of the temperature in the HCS. Consequently, the ''new''
collision operator $J^{^{\prime }}(f)$on the right side has properties
similar to that for elastic collisions: an invariant state and $5$ vanishing
low velocity moments 
\begin{equation}
J^{^{\prime }}(F^{(0)})=0,\,\quad \int d{\bf v}\psi _{\alpha }\left( {\bf v}%
\right) J^{^{\prime }}(f)=0,\hspace{0.5cm}\psi _{\alpha }\longleftrightarrow
(1,m{\bf v,}\frac{1}{2}mV^{2}).  \label{7.3}
\end{equation}%
Now the usual BGK kinetic model for elastic collisions can be used to
represent $J^{^{\prime }}(f)$ rather than $J(f)$, preserving the properties (%
\ref{7.3}) 
\begin{equation}
J^{^{\prime }}(f)\rightarrow -\omega \left( f-F^{(0)}\right) .  \label{7.4}
\end{equation}%
where $F^{(0)}$ is the local HCS distribution. The collision frequency $%
\omega $ has a space and time dependence that occurs only through the
density and temperature, and is a free parameter of the model. With this
choice the model kinetic equation becomes 
\begin{equation}
\left( \partial _{t}+{\bf v\cdot \nabla }\right) f=-\omega \left(
f-F^{(0)}\right) +\frac{1}{2}\zeta \frac{\partial }{\partial {\bf v}}\cdot
\left( {\bf V}f\right) .  \label{7.5}
\end{equation}%
A review of some applications of kinetic models for granular gases is given
in reference \cite{brey00model}. Such methods are potentially more important
for granular gases since the states of interest are typically driven by
external boundary conditions, posing intractable difficulties for analytic
studies of the Enskog equation.

{Exact solutions in kinetic theory for spatially inhomogeneous states are
exceedingly rare. When, furthermore, such a solution is normal and
corresponds to a hydrodynamic state far from equilibrium an important
benchmark is obtained for both conceptual and computational issues. In the
case of granular gases at low density there are two interesting examples:
uniform shear flow \cite{brey97shear} and planar nonlinear Couette flow \cite%
{tij01}. Both allow controlled discussion of nonlinear rheological
properties (e.g., shear thinning, normal stresses) that are important for a
wide class of real granular flows. For uniform shear flow the model results
predict the components of the \ pressure tensor as a function of shear rate
and }$\alpha $, as well as the velocity distribution function itself which
is manifestly normal. For nonlinear Couette flow the hydrodynamic fields,
heat and momentum fluxes, and velocity distribution function are determined
for conditions such that viscous heating dominates collisional cooling,
including large gradients. Explicit expressions for the generalized
transport coefficients (e.g., viscosity and thermal conductivity) are
obtained as nonlinear functions of the shear rate {and }$\alpha $. {%
Comparison of such non Newtonian transport properties with DSMC results for
the Boltzmann equation in each case shows surprisingly good agreement,
confirming that the simplifications of the model do not compromise the
physical implications of the Boltzmann equation. For the present discussion,
these examples demonstrate the existence of a normal solution and
hydrodynamics under conditions for which the Chapman-Enskog expansion is no
longer justified.}

\section{Separation of scales}

It remains to discuss the conditions for dominance of a hydrodynamic
description. Simply put, the time scale for hydrodynamic processes must be
long compared to that required for all other (''kinetic'') excitations to
decay and become negligible. The characteristic longest time scale for the
kinetic excitations is of the order of the mean free time $\tau _{m}$. The
analysis is simplest for states near the homogeneous state (equilibrium or
HCS). Then the amplitudes of the hydrodynamic fields do not deviate much
from their uniform values, but their spatial variation can be complex. For
fluids with elastic collisions the hydrodynamic frequencies, $\omega _{H}(k)$%
, are functions of the wavevector $k$ (inverse wavelength) such that $\omega
_{H}(k)\rightarrow 0$ as $k\rightarrow 0$. Consequently, a sufficient
condition for the separation of time scales is a restriction to sufficiently
small $k<\ell ^{-1}=$ (mean free path)$^{-1}$, or in terms of the frequency $%
\omega _{H}(k)<\omega _{0}=v_{0}\ell ^{-1}=\tau _{m}^{-1}$ collision
frequency. Note that this does not require that the mean free path should be
large compared to the particle size. For example, in dense fluids the mean
free path can be smaller than the hard sphere diameter and an accurate (but
not Navier-Stokes) hydrodynamics applies from macroscopic length scales down
to those of the order of the particle size. The analysis for inelastic
collisions is similar, except that there is the additional condition that
the cooling rate must be smaller than the collision rate, $\zeta <\omega
_{0} $, in order to include the temperature as one of the hydrodynamic
variables.

For states near the HCS, the separation of time scales (or frequencies) for
a granular gas can be studied from the spectrum of the associated linear
kinetic equation. There are two cases for which precise and unambiguous
results have been obtained: the Enskog-Lorentz equation for a heavy impurity
particle diffusing through a fluid in its HCS, and a kinetic model for the
Boltzmann equation. In the first case a mass ratio expansion of the
Enskog-Lorentz equation leads to a Fokker-Planck equation \cite%
{brey99model3,santos01}. While there are many qualitative differences from
the corresponding equation for elastic collisions, there is an exact mapping
of the case $\alpha <0$ to that with $\alpha =1$ for all values of the
impurity-gas restitution coefficient. The spectrum of the Fokker-Planck
equation can be obtained exactly in this way. It shows a diffusive mode
smaller than infinitely many other discrete kinetic modes, bounded away from
them at a smallest distance of the order of the impurity-gas collision
frequency. Furthermore this distance remains the same independent of the
wavelength considered. Thus the justification and conceptual basis for \
hydrodynamic diffusion in this granular ''Brownian motion'' is the same as
for elastic collisions; it dominates after a few collisions. The predictions
of this Fokker-Planck description \ have been confirmed in detail by DSMC
for the Boltzmann-Lorentz equation and by MD \cite{brey99brown}.

Similar conclusions follow for the linearized kinetic model for the
Boltzmann equation (\ref{7.5}). As an example, consider an initial state
that excites transverse shear excitations (a perturbation of the transverse
flow field) \cite{brey99model3}. The complete spectrum can be determined in
this case as well, although it is more complex including branch cuts as well
as discrete points. There is an isolated shear diffusion mode with
dimensionless frequency%
\begin{eqnarray}
\omega _{H}^{\ast }\left( k,\alpha \right) &=&\frac{\zeta ^{\ast }}{2}%
-\left( 1+\frac{\zeta ^{\ast }}{2}\right) \omega _{H}^{\ast }\left( \frac{%
k^{\ast }}{1+\frac{\zeta ^{\ast }}{2}},\alpha =1\right)  \nonumber \\
&\rightarrow &\frac{\zeta ^{\ast }}{2}-\frac{k^{\ast 2}}{2+\zeta ^{\ast }}
\label{8.1}
\end{eqnarray}%
where $\zeta ^{\ast }=\zeta /\omega _{0}$ is the dimensionless cooling rate.
The second line indicates the small wavevector limit of $\omega _{H}^{\ast
}\left( k^{\ast },\alpha =1\right) $. This is the unstable shear mode
discussed in Section 4. In this case it is extended to larger wavevectors,
beyond the Navier-Stokes form in the second line, but there is a restriction
on the wavevector to be less than the inverse mean free path (more precisely 
$k^{\ast }=k\ell \leq $ $\left( 1+\frac{\zeta ^{\ast }}{2}\right) \sqrt{\pi }
$ ), just as for elastic collisions. The remainder of the spectrum is at
larger frequencies with a smallest separation of the order of the collision
frequency. Hence shear diffusion dominates after a few collisions. The
longitudinal excitations can be studied exactly in the same way for this
kinetic model, with the same conclusion that the hydrodynamic spectrum
remains bounded away from the kinetic modes and sufficiently small
wavevectors.

In both cases the analysis holds without any restriction to weak
dissipation, and the cooling rate always remains separated from the kinetic
modes. This provides an illustration of a contracted description including
the temperature as one of the variables, even though it is not associated
with a conserved quantity. Within the set of hydrodynamic excitations there
can be a second separation of time scales when $k^{\ast 2}<<\zeta ^{\ast }$,
i.e. long wavelengths and strong dissipation. This is possible since some of
the hydrodynamic modes vanish as $k^{\ast 2}\rightarrow 0$ while others are
coupled to the temperature whose dominant time scale $\zeta ^{\ast -1}$ is
fixed by $\alpha $. Calculations based on the Boltzmann and Enskog equations
lead to the necessary condition for the temperature to be a hydrodynamic
field, $\zeta ^{\ast }<\omega _{0}$. Then hydrodynamics includes the two
qualitatively different regimes, $\zeta ^{\ast }<<k^{\ast 2}<\omega _{0}$
and $k^{\ast 2}<<\zeta ^{\ast }<\omega _{0}$. Occurrence of this separation
within the hydrodynamic domain should not be confused with the last
inequality, which assures the hydrodynamic description. The inclusion of a
non-conserved degree of freedom, temperature, is not unique to granular
gases. Some complex molecular gases have internal molecular modes with
frequencies comparable to the sound velocity, and these modes can be
considered as ''hydrodynamic'' on the dominant long time scale.

The above analysis refers to states near the HCS, confirming the {\em %
sufficient} conditions of hydrodynamic fields whose space and time variation
is smooth on the scale of the mean free time and mean free path. When the
hydrodynamics is non Newtonian, the field variations are large and these
sufficient conditions are necessarily violated. Nevertheless, it is possible
that the characteristic hydrodynamic frequencies can still be small compared
to the mean free path and the separation of time scales can still be valid.
This is the case for polymer fluids, where rheological properties imply
shear rates large compared to the collision frequency. It is difficult to
provide a general characterization of non Newtonian hydrodynamics since the
constitutive equations are not known and may not be universal. However,
uniform shear flow is a rare case for which exact results can be obtained %
\cite{shear1} to illustrate the existence of hydrodynamics for large spatial
gradients. An exact, closed set of equations for the components of the
pressure tensor can be obtained from the Boltzmann equation for the special
case of Maxwell molecules. The solution shows an exponentially fast approach
to the hydrodynamic state (uniform shear flow with viscous heating, as
obtained from the hydrodynamic equations) for times large compared to $\tau
\left( a^{\ast }\right) $ 
\begin{equation}
\tau \left( a^{\ast }\right) =\tau _{m}\left[ 2\left( 1+9a^{\ast 2}+3a^{\ast
}\sqrt{2+9a^{\ast 2}}\right) ^{1/3}+2\left( 1+9a^{\ast 2}-3a^{\ast }\sqrt{%
2+9a^{\ast 2}}\right) ^{1/3}-3\right] ^{-1}  \label{8.2}
\end{equation}%
where $a^{\ast }=a\tau _{m}$ is the shear rate times the mean free time,
measuring the hydrodynamic gradient. It is seen that $\tau \left( a^{\ast
}\right) \leq \tau _{m}$, for {\em all} values of $a^{\ast }$, so the
hydrodynamic description applies as expected for $t\gg \tau _{m}$ even if $%
a^{\ast }>>1$. A corresponding approximate analysis for hard spheres based
on a kinetic model can be extended to inelastic collisions \cite{brey97shear}%
, leading to similar conclusions. The steady shear flow of a granular gas
necessarily entails large $a^{\ast }$ (see comments below) and this has been
interpreted as a breakdown of hydrodynamics for granular systems \cite{tan98}%
. However, as the above example illustrates, it is only a failure of
conditions for the {\em Navier-Stokes} form of hydrodynamics, not
hydrodynamics more generally \cite{dufty99comment}.

It is appropriate at this point to remark on the conjecture by Goldhirsch in
his recent review \cite{goldhirsh01}: ''The main thesis of this paper is
that granular gases should be considered to be mesoscopic in the sense that
both the microscopic spatial and temporal scales are typically not well
separated from the relevant corresponding macroscopic scales and this
property of granular gases is at the root cause of many and perhaps most of
the peculiar properties of granular gases.'' The terminology ''not well
separated'' requires analysis in each case. It is not sufficient to make
estimates based on dimensional analysis of macroscopic fields in terms of
characteristic mesoscopic (kinetic) scales. As the example of the last
paragraph shows, such analysis can lead be misleading (i.e. $a^{\ast }>>1$
does not imply the time scales are not well separated). The test for the
validity of a hydrodynamic description for granular gases must be more
precise: how do the true hydrodynamic frequencies compare to the true
kinetic frequencies? For granular gases both the physical states and the
forms for these frequencies can be more complex than those for normal
fluids. The failure of familiar properties for the latter to hold for the
former does not necessarily imply a failure of hydrodynamics, and there
appears to be no precise example of hydrodynamic frequencies exceeding the
collision frequency. Still, Goldhirsh is correct to warn that granular
systems frequently support states outside the Navier-Stokes domain and
therefore properties based on the latter cannot be trusted.

It is argued in \cite{goldhirsh01} that mean free paths in granular systems
can be macroscopic, leading to a failure to separate microscopic and
macroscopic space scales. The primary example is a mean free path defined in
the laboratory frame rather than the local rest frame for the fluid element.
Such a definition allows arbitrarily large mean free paths for sufficiently
large convection of the fluid, and can be anomalously large for both normal
and granular fluids. Reasons for rejecting this definition are given in \cite%
{dufty99comment}. In general, when properly formulated it appears that the
mean free path in a granular medium is qualitatively similar to that in a
normal fluid. Of course, in strongly heterogeneous states of dense clusters
and large voids the average mean free path is no longer a useful concept.

The title of the review \cite{goldhirsh01} ''Granular gases: probing the
boundaries of hydrodynamics'' recognizes an important motivation for
studying rapid granular flow. The hydrodynamics is indeed different from the
standard Navier-Stokes form for many reasons, and commonly occurring states
often sample a different domain of the relevant parameter space than is
accessible for\ normal fluids. The single feature of an energy sink allows a
wide range of phenomena that are not possible otherwise. For example, the
steady states of uniform shear flow or Couette flow described in Section 7
are possible only because of the balance between viscous heating and
collisional cooling. In the steady state, therefore, the temperature, shear
rate $a$, and restitution coefficient $\alpha $ are no longer independent
variables. In dimensionless terms the relationship gives $a^{\ast }\propto
\left( 1-\alpha ^{2}\right) $ and it is possible to show that the fluid is
always non Newtonian for any $\alpha <1$ \cite{montanero01viscosity}. A
similar conclusion applies for a granular gas confined between walls with
fixed temperature. The resulting steady state has a nonuniform temperature
profile resulting from competition between the energy source at the wall and
collisional cooling in the \ interior. The profile is fixed by the
restitution coefficient and wall temperatures. Again, it can be shown that a
correct description of the profile cannot be obtained from the Navier-Stokes
equations due to inherent non Newtonian effects induced by the steady \
state. This is probably a characteristic of most steady states obtained by
driving the system. Such non Newtonian behavior of simple atomic systems is
impossible to observe experimentally (except in computer simulations) but
seems to be the norm for granular gases. This makes the search for complex
constitutive equations an interesting and challenging research area to be
pursued. Rather than abandoning hydrodynamics for granular gases, it appears
profitable to embrace such a description for new horizons and new
opportunities.

\section{Boundary value problems}

Most problems of practical interest involve finite geometries and associated
boundary conditions. Such problems can be addressed directly at the level of
the Navier-Stokes hydrodynamics but, as with normal fluids, become less
straightforward for non Newtonian flows. In any case, boundary value
problems do not appear to pose any inherent problem for kinetic theory or
its implementation via DSMC. As noted by Grad in his discussion of the
Boltzmann equation \cite{grad58} a normal solution of the form \ref{5.10}
can be expected to apply only outside certain domains of initial and
boundary ''slip'' (a third slip across a shock layer is also discussed). The
initial slip layer is the transient period of a few collisions required for
the kinetic excitations to decay relative to those of hydrodynamics, as
described above. Similarly, there is a domain of the order of a few mean
free paths near the boundaries for which the solution to the kinetic
equation is more complex than that which can be represented by the normal
form. Away from these slip domains a description in terms of the five
hydrodynamic fields can be expected. The boundary conditions for these
fields necessarily require an independent means for extrapolation across the
slip domains (modified initial and boundary conditions, relative to the
actual given conditions).

For large systems and Navier-Stokes order gradients, the slip conditions are
often negligible and the interesting physical phenomena resulting from
boundary sources are not dependent on the mesoscopic details of the boundary
layer. Recently, Brey et al. \cite{brey01bound} have applied the
Navier-Stokes hydrodynamic equations derived from the Boltzmann equation as
described in Section 6 to describe a vibrated granular gas in an external
gravitational field. It is worth repeating that the transport coefficients
and cooling \ rate are specified functions of the restitution coefficient so
there are no adjustable parameters. The system was driven at $z=0$ and
unbounded for $z>0$. In appropriate dimensionless variables, non-trivial
temperature and density profiles are obtained independent of the details of
the boundary conditions (frequency and amplitude of vibration) although a
transition to a collisionless Knudsen gas eventually occurs at
asymptotically large $z$. These predictions from Navier-Stokes hydrodynamics
are in very good agreement with corresponding results from both MD and DSMC
for the Boltzmann equation. Similar good agreement between Navier-Stokes
hydrodynamics, MD, DSMC, and experiment has been observed for gravitational
flow of grains past a fixed wedge \cite{rericha01}. In this case the flow is
supersonic leading to formation of a shock profile at the tip of the wedge
and an expansion fan below the wedge, all well-described by the hydrodynamic
boundary value problem.

It is interesting to note that the Navier-Stokes equations are capable of
describing some of the more exotic behavior exhibited by granular gases,
even though structurally they are quite similar to the equations for normal
fluids. Symmetry breaking is one such phenomenon. Consider the above shaken
system now partitioned vertically into two domains initially with the same
density profiles. If a hole in the partition is introduced at a certain
height, an asymmetry of the density in the two partitions develops depending
on the amplitude of the vibration and degree of dissipation. The description
in terms of a simple effusion process is in reasonable agreement with
results from experiment and MD \cite{eggers99}. A more detailed
Navier-Stokes analysis of a different symmetry breaking mechanism also is in
good agreement with MD \cite{brey01demon}. \ In this case a closed container
of $N$ particles is vibrated at $z=0$ (no gravity) and an exact solution to
the Navier-Stokes equations with constant pressure is constructed. Next, the
system is partitioned normal to $z$ starting at a height $z=h$. At
sufficiently low density the hydrodynamic fields are symmetric with respect
to the two sides of the partition, but above a critical density an asymmetry
in the density occurs (a bifurcation in the hydrodynamic solution). The
asymmetry (high and low density sides) continues to increase with larger
density. The hydrodynamic analysis again is in excellent quantitative
agreement with MD and DSMC results.

If some dimensions of the system are small, the details of the boundary
conditions can become more important. For granular systems the boundaries
may be rough, e.g. consisting of frozen layers of other grains, and the
geometry of this roughness (local curvature) also can play a role. Coulomb
friction is another mechanism for momentum transfer at the boundary. The
problem of constructing realistic boundary conditions for granular gas
hydrodynamics has been reviewed and illustrated recently by Jenkins \cite%
{jenkins01}. As in the case of normal fluids, the objective is not to
describe in detail the boundary layer near the surface, but rather to
construct representative conditions for the hydrodynamic equations on the
other side of the slip. Practical boundary conditions have been found in
many cases for the accurate application of hydrodynamics for experimental
analysis. This is an active and evolving field.

\section{Discussion}

Two questions have been addressed here: 1) can kinetic theory provide a
valid mesoscopic description of rapid flow (fluidized) granular media?, and
2) can a hydrodynamic description be formulated and justified for a
macroscopic description? The evidence presented here has emphasized
idealized conditions and states for which controlled numerical simulations
and theoretical approximations can be compared. At this level, it appears
that many of the questions, concerns, and objections raised regarding the
kinetic theory\ have been removed and the domain of validity has been
clarified in the past few years. Certainly, kinetic theory (Enskog or more
sophisticated) appears to be a powerful tool for analysis and predictions of
rapid flow gas dynamics. Combined with the numerical solution via DSMC,
almost any boundary or initial value problem for moderate densities can be
explored in detail. The most practical version of kinetic theory, the Enskog
equation, appears to have a somewhat more restricted domain in the parameter
space of density and restitution coefficient than for fluids with elastic
collisions, but remains a remarkably rich and accurate basis for analysis of
moderately dense granular gases. Any observable differences from Enskog
provide a new opportunity to study correlated collision phenomena whose
effects for normal fluids are typically small except for the dense fluid.

The hydrodynamic description for the same class of idealized conditions and
states, also appears justified even though the phenomena can be considerably
more complex than for normal fluids. Nevertheless, this complexity (e.g.,
instabilities, clustering, \ rheology) appears to be captured quantitatively
by a properly formulated hydrodynamics. The required separation of time
scales, including the temperature field, appears justified even at strong
dissipation based on the few cases for which this question can be addressed
quantitatively. For weakly inhomogeneous states the Chapman-Enskog method,
properly implemented, gives both the Navier-Stokes hydrodynamics for a
granular gas and the detailed forms for the transport coefficients as
functions of density and $\alpha $. Quantitative confirmation of these
results, as indicated above in many studies, should remove most reservations
about the basis of hydrodynamics for such states. Non Newtonian
hydrodynamics for states in which the gradients of hydrodynamic fields are
not small is less well-understood, just as for normal gases. However, in the
latter case such states are rare or unphysical for simple atomic systems. In
contrast, they appear frequently for granular gases in steady states where
the gradients are strongly correlated to the coefficient of restitution.
Kinetic models and DSMC for shear and Couette flow provide instructive
examples of non Newtonian hydrodynamics, and support for the expectation
that hydrodynamics with appropriate constitutive equations is applicable to
such complex rheological states as well.

The optimistic view about hydrodynamics for granular media presented here
appears at odds with the conclusion by Kadanoff in \cite{kadanoff99}: ''Can
a granular material be described by hydrodynamic equations, most
specifically those equations which apply to an ordinary fluid? It seems to
me the answer is ''No!''''. This conclusion is based on a number of observed
phenomena: weakly shaken compact material showing composite effects and
patterns, compaction (volume reduction) of dense material by tapping,
clustering as described in Section IV, inelastic collapse (see footnote in
Section 2), and one dimensional examples. It is important to note first that
Kadanoff has in mind Navier-Stokes-like hydrodynamics (universal, local,
partial differential equations). Indeed, in many of these cases it is
difficult to imagine a hydrodynamic description of the type requested. These
are not the fully fluidized states considered here, and it is clear that the
presence of compact structures or large voids entail heterogenous materials
for which Navier-Stokes hydrodynamics was never intended. The case for
hydrodynamics improves considerably when that term is extended to include
more general constitutive equations and when the states considered are fully
fluidized in two or three dimensions. Polymer and colloidal fluids, liquid
crystals, and emulsions all admit a hydrodynamic descriptions although more
complex than Navier-Stokes, which are neither universal nor necessarily
local. A single granular material can manifest properties similar to many
different states of complex fluids, and it is possible that many different
forms of constitutive equations will \ be required for an adequate
hydrodynamic description of each. A single normal material also can have
multiple types of macroscopic continuum descriptions for different states
(fluid, metastable, elastic). One conclusion here is that granular media
exhibit a wide range of interesting phenomena for which a Navier-Stokes
hydrodynamics is an accurate and practical tool. More generally, however,
there is agreement with Kadanoff that granular fluids are in a class of
complex materials ''with behaviors which are, at this moment, not fully
understood.'' Kinetic theory and hydrodynamics (in the broader sense) can be
expected to provide much of this understanding.

\section{Acknowledgements}

It is a pleasure to acknowledge the essential assistance of Javier Brey,
Vicente Garz\'{o}, James Lutsko, and Andr\'{e}s Santos in their roles as
collaborators, mentors, and critics for much of the material discussed here.
This research was supported in part by National Science Foundation grant PHY
9722133.

\end{document}